\documentclass[twoside]{article}
\usepackage{fleqn,espcrc2}
\usepackage{graphicx}
\usepackage{amsmath}
\usepackage{epsfig}

\title{NNPDF1.0 parton set for the LHC}

\author{ M.~Ubiali \address{ Department of Physics, University of Edinburgh, Edinburgh, UK} \thanks{M.Ubiali is founded by a SUPA studentship.}}

\begin{document}

\begin{abstract}
We present the first NNPDF full set of Parton Distribution Functions from a comprehensive DIS analysis. This approach, combining a Monte Carlo sampling of the probability measure in the space of PDFs with the use of neural networks as interpolating functions, provides a faithful and statistically sound determination of the uncertainty in parton distributions. The features of the fit and the results are discussed in details as well as some preliminary phenomenological analysis.\end{abstract}

\maketitle

\section{INTRODUCTION}

One of the key elements for the computation of any observable involving hadrons are Parton Distribution Functions (PDFs); a faithful estimation of their error is fundamental for producing reliable phenomenological predictions at hadronic colliders. Especially now, with the upcoming LHC data and with increasingly smaller experimental uncertainties, a careful consideration has to be given to theoretical uncertainties. 

The traditional approach for PDFs fitting\cite{MRST,CTEQ,alekhin2} suffers of some drawbacks which have not been completely solved. In particular, benchmark comparison performed between some of those sets~\cite{heralhc} shows that benchmark partons determined on restricted data sets and global fit partons do not agree within error. This makes the uncertainty bands not easily interpretable in a statistical sense. This and other difficulties have stimulated various proposals for new approaches. 

The NNPDF approach is one of them. It has been introduced in the context of the parametrisation of DIS structure function data~\cite{f2ns,f2p} and, after having been successfully applied in the determination of the non-singlet distribution~\cite{nnqns}, it has been used for the construction of a full partonic set from DIS data~\cite{nnsinglet}. The details of the methodology and analysis are widely explained in Ref~\cite{nnsinglet}. In the following section we briefly describe the method and the features of the fit by concentrating especially on the results. 

\section{THE NNPDF1.0 PARTON SET}

\begin{figure}[htb]
\begin{center}
\vspace{-0.7cm}
\includegraphics[width=7cm]{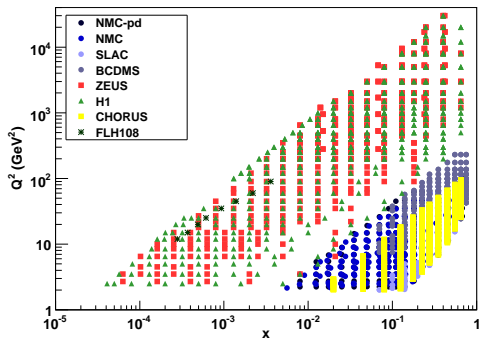}
\vspace{-0.3cm}
\end{center}
\caption{Experimental data used in the analysis after kinematical cuts.}
\label{fig:kin}
\end{figure}

The determination of the NNPDF1.0 set is based on a full set of deep-inelastic scattering data, with various lepton beams and nucleon targets for a total of more than 3000 measurements coming from 7 different experiments. The kinematic coverage of data sets included in the analysis are shown in Fig.~\ref{fig:kin}. The observables used in our fit are either structure functions or reduced cross-sections, including neutrino reduced cross section.

In order to propagate the error from the experimental data to the fit we build a sampling of the probability distribution defined by the experimental data. To do this, we generate $N_{\mathrm{rep}}$ sets of artificial data distributed according to a multi-gaussian distribution centered on the original data and with variance determined by experimental uncertainties. The accuracy of the statistical sampling is quantified by mean of statistical estimators which indicate that a sample of ${\mathcal O}(1000)$ replicas is enough to reproduce the mean values, the variance and the correlations of the experimental data within a 1\% accuracy.

For each replica we fit the artificial data by evolving the PDFs from the starting scale to the scale of the experimental points. Each of the five input PDFs is parameterized by a multi-layer neural network. The latter provide nothing but a redundant and unbiased parameterization for the PDFs at the initial scale. In principle, any other redundant parametrisation with the same features would be suitable.

The determination of the best fit in case of a redundant parameterization is a delicate issue because it might adapt not only to the physical behaviour but also to statistical fluctuations. Therefore the best fit is given by an optimal training, beyond which the figure of merit improves by learning the statistical noise of the data. We address this issue through a so called cross-validation method, based on a random division of data into a training and a validation set. The first set is the one on which we actually minimize the figure of merit, which is a function of the weights of the nets.
The latter is evaluated not only on the training set but also over the validation set at each iteration of the minimisation, as a monitor. In fact, when the error function of the training set still decreases and the validation one starts increasing we have reached the optimal fit.
It is extremely important that the best fit is not determined as the absolute minimum for a given functional form; in this way inconsistent data or underestimated uncertainties are automatically accounted for and signalled by a larger than average value of the $\chi^2$ per degree of freedom and do not require a separate treatment.

\begin{figure}[htb]
\vspace{-0.3cm}
\includegraphics[width=6.5cm]{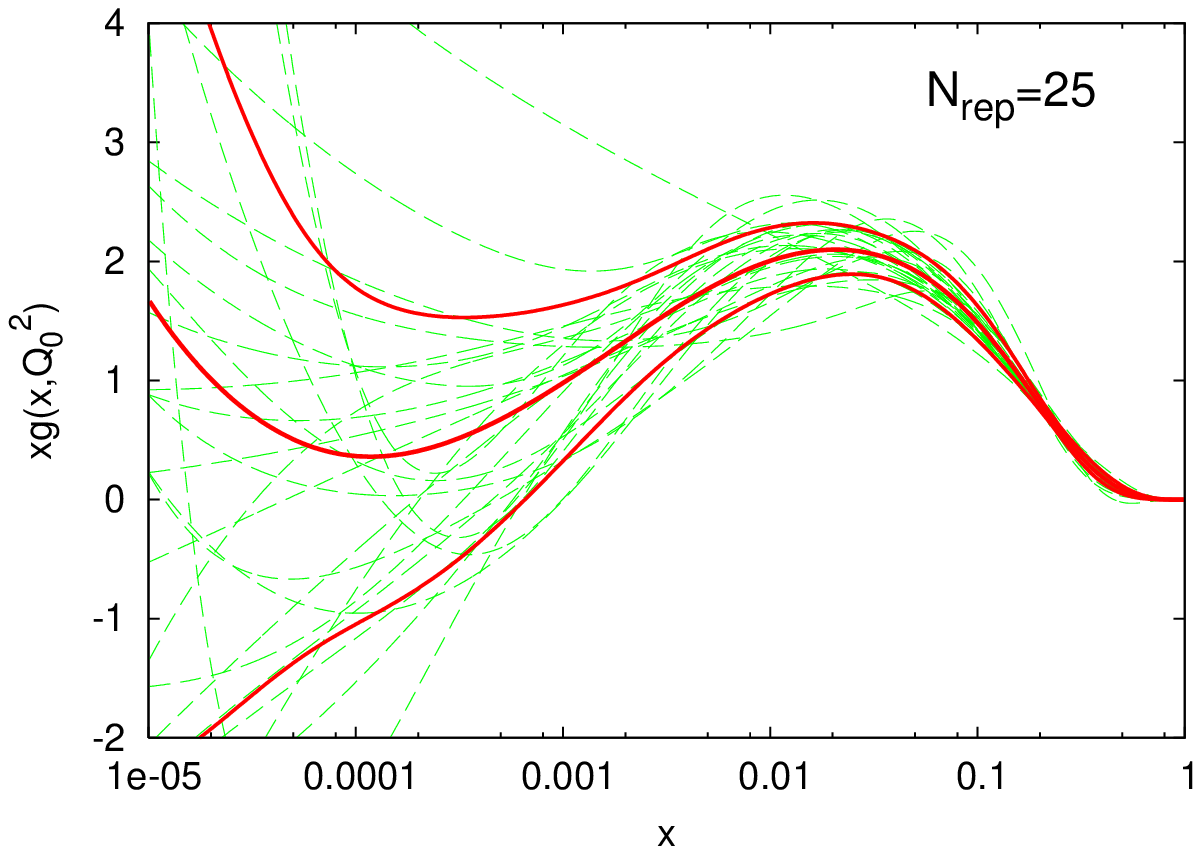}
\includegraphics[width=6.5cm]{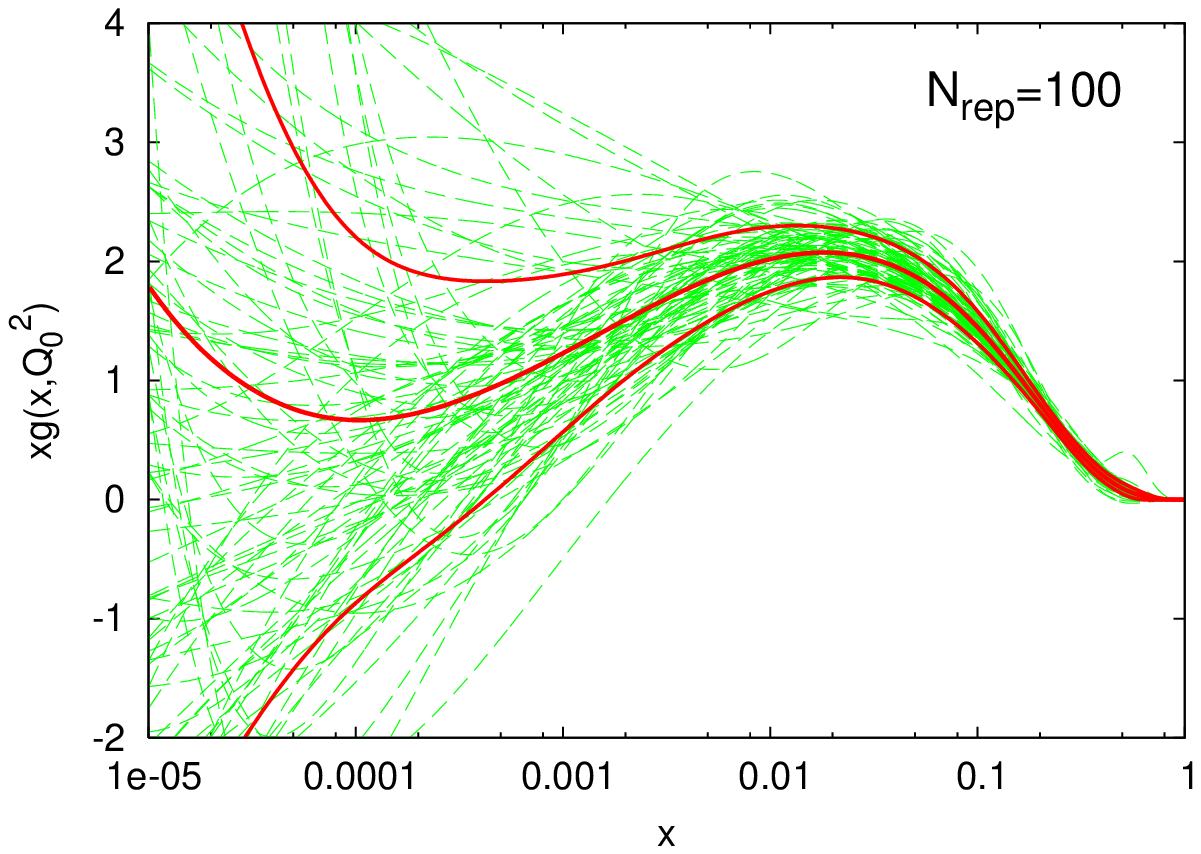}
\caption{Set of 25 replicas (top) and 100 replicas (bottom) of the gluon distribution. The solid red line show the central value and the one-sigma interval computed from each set.}
\label{fig:osc}
\end{figure}

At the end of this process we end up with a set of $N_{\mathrm{rep}}$ trained neural networks which provides a representation of the probability density. In Fig.~\ref{fig:osc} it is shown that, even though individual replicas may fluctuate significantly because of the flexibility of the parameterization, average quantities such as central values and error bands are smooth inasmuch as stability is reached due to the number of replicas increasing.
Therefore any statistical property of the parton distributions themselves or of any function of them can be calculated using standard statistical methods. It is thus easy to compute any desired property such as correlations or to assess the stability of the fit under the variation of the number of parameters describing the basis PDFs. It is also easy to restrict our fit on a subset of data and verify that, while the uncertainty bands do increase in the region where there are less data, the central values remain compatible within uncertainty~\cite{nnsinglet}. 

With the method briefly described we have produced the NNPDF1.0 set of parton distributions. In our analysis we fit five independent PDFs corresponding to the two light flavors and the gluon. We assume the strange distribution to be proportional to the light sea given that the data sets included in the analysis give little constraints on the strange PDFs. Besides, we determine all heavy quark PDFs from perturbative evolution, generating them dynamically according to the ZM-VFN scheme and therefore neglecting intrinsic heavy flavor contributions. Evolution is performed at next-lo-leading order from the initial scale $Q_0^2=$2 GeV$^2$=m$_c^2$.
The quality of the central fit, obtained averaging over all PDFs in the sample is measured by its $\chi^2=1.34$. 
\begin{figure}[htb]
\vspace{-0.7cm}
\includegraphics[width=6.5cm]{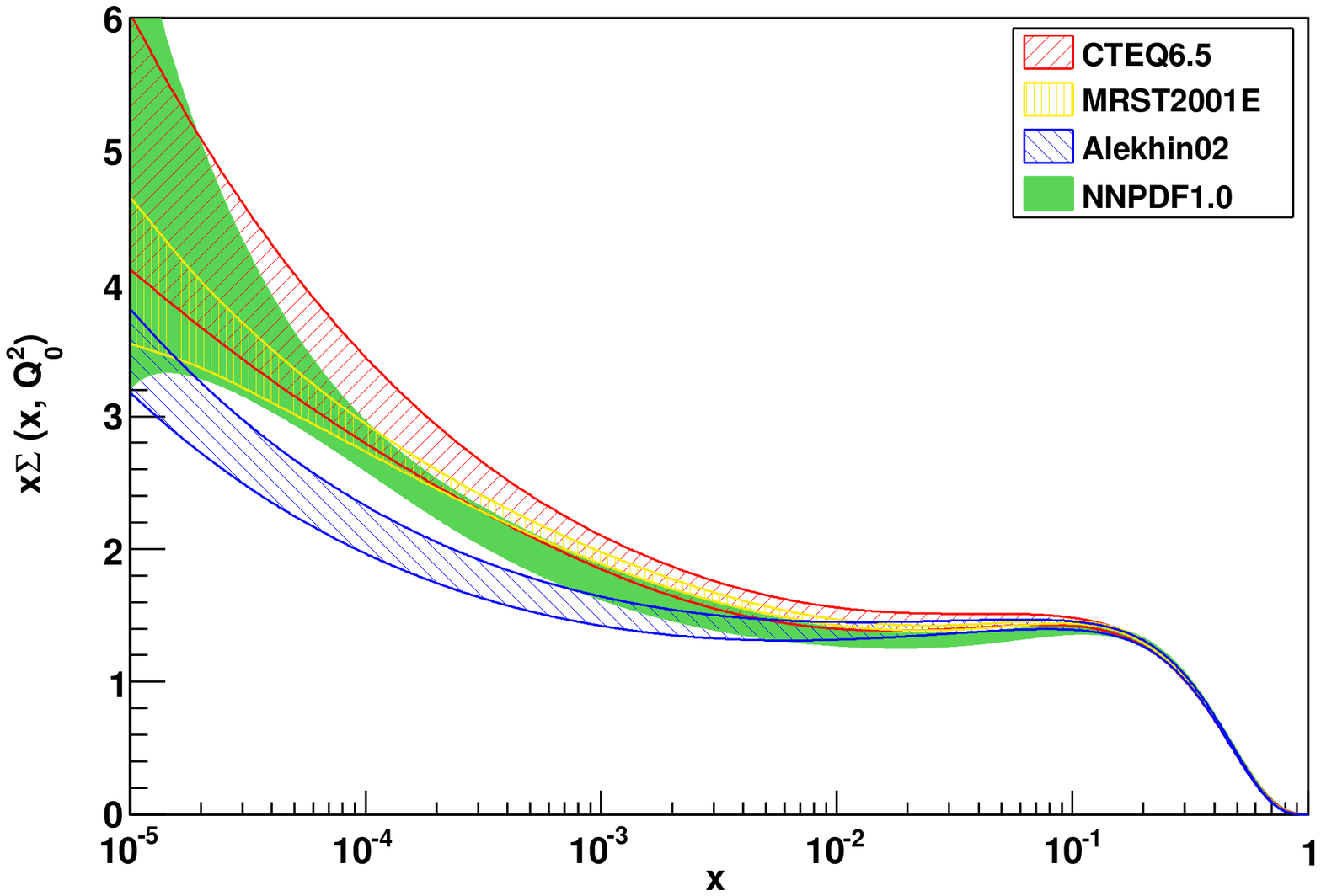}
\includegraphics[width=6.5cm]{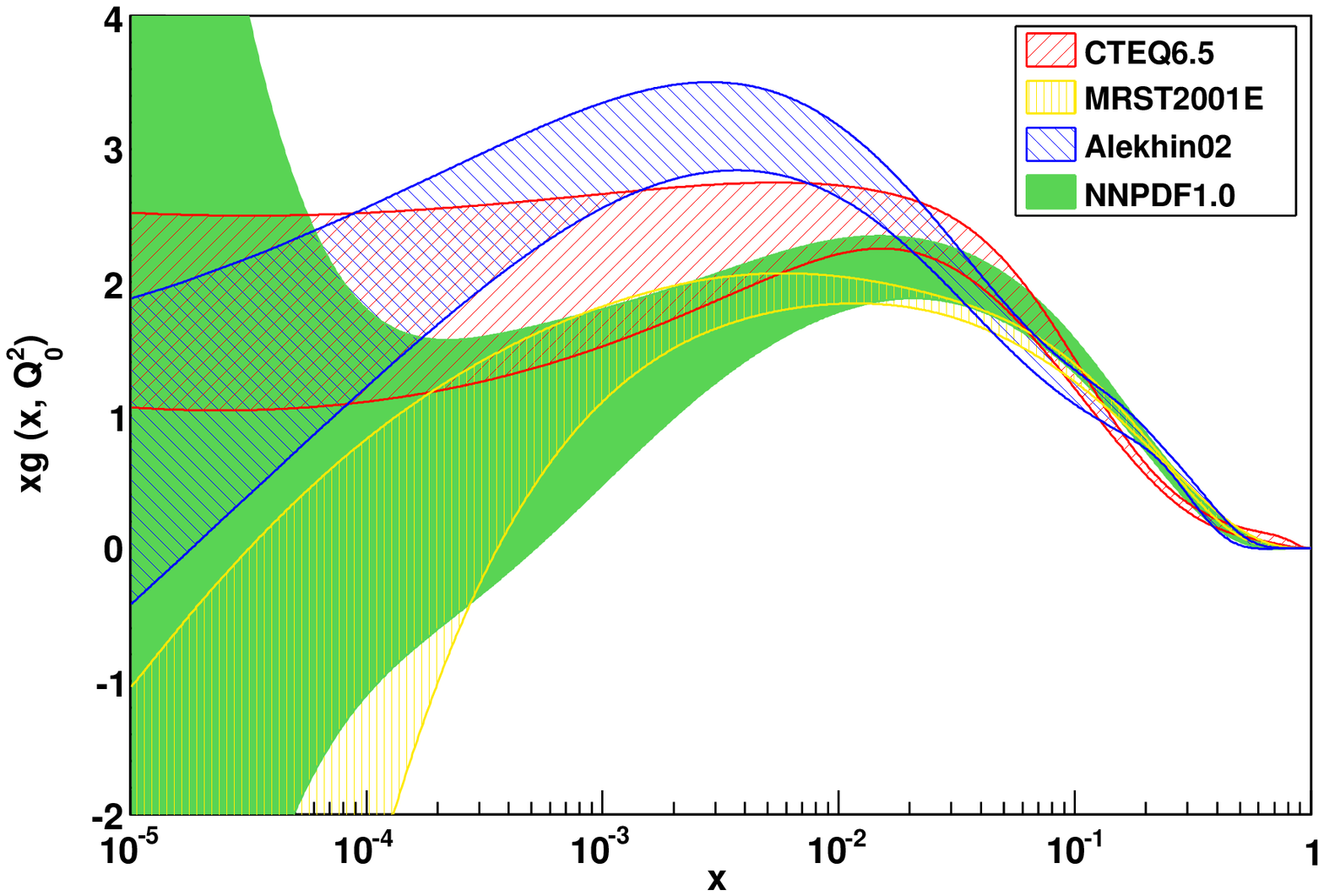}
\caption{The singlet and gluon PDF at the initial scale $Q_0^2=2$ GeV$^2$.}
\label{fig:sg}
\end{figure}

\begin{figure}[htb]
\vspace{-0.3cm}
\includegraphics[width=6cm]{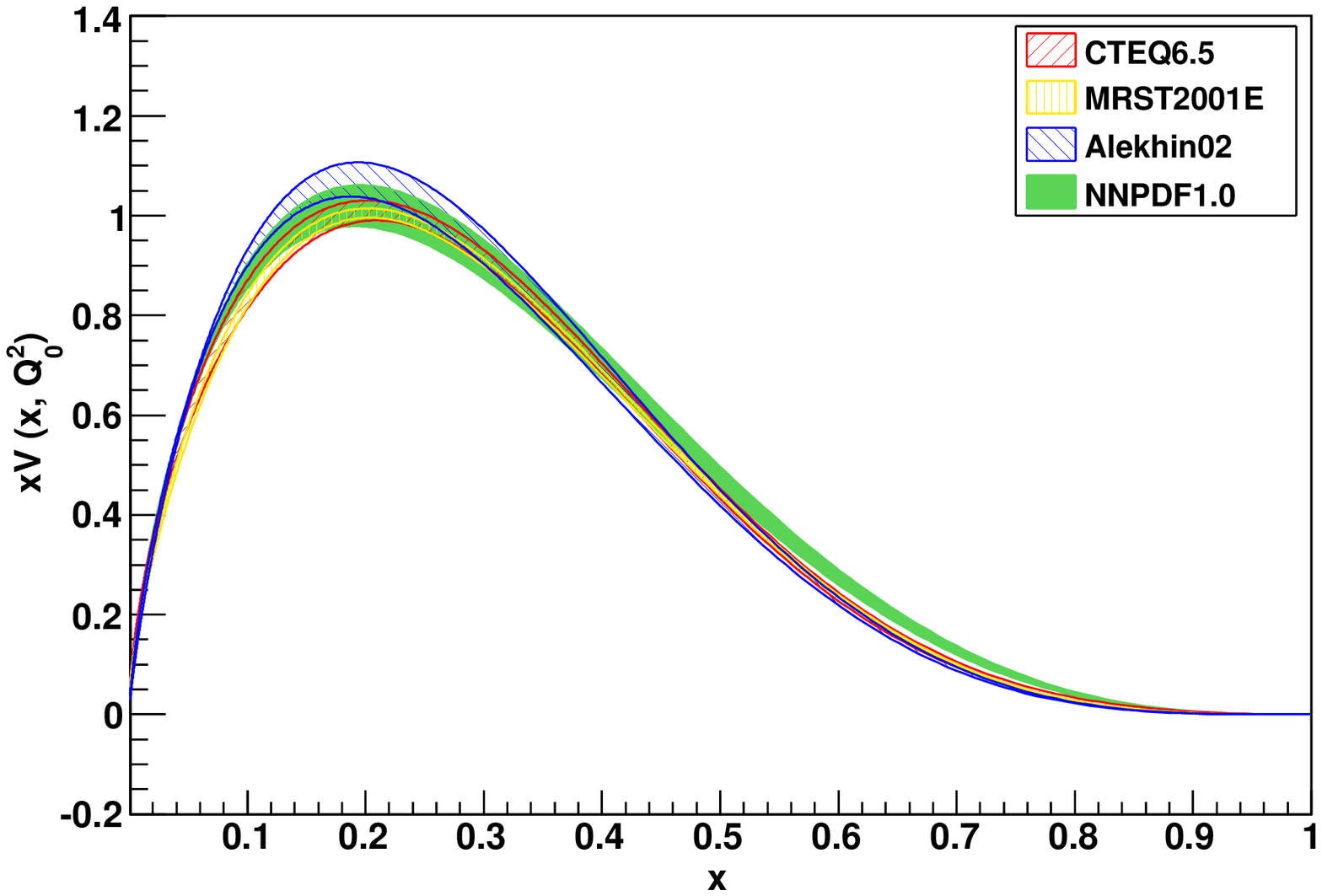}
\includegraphics[width=6cm]{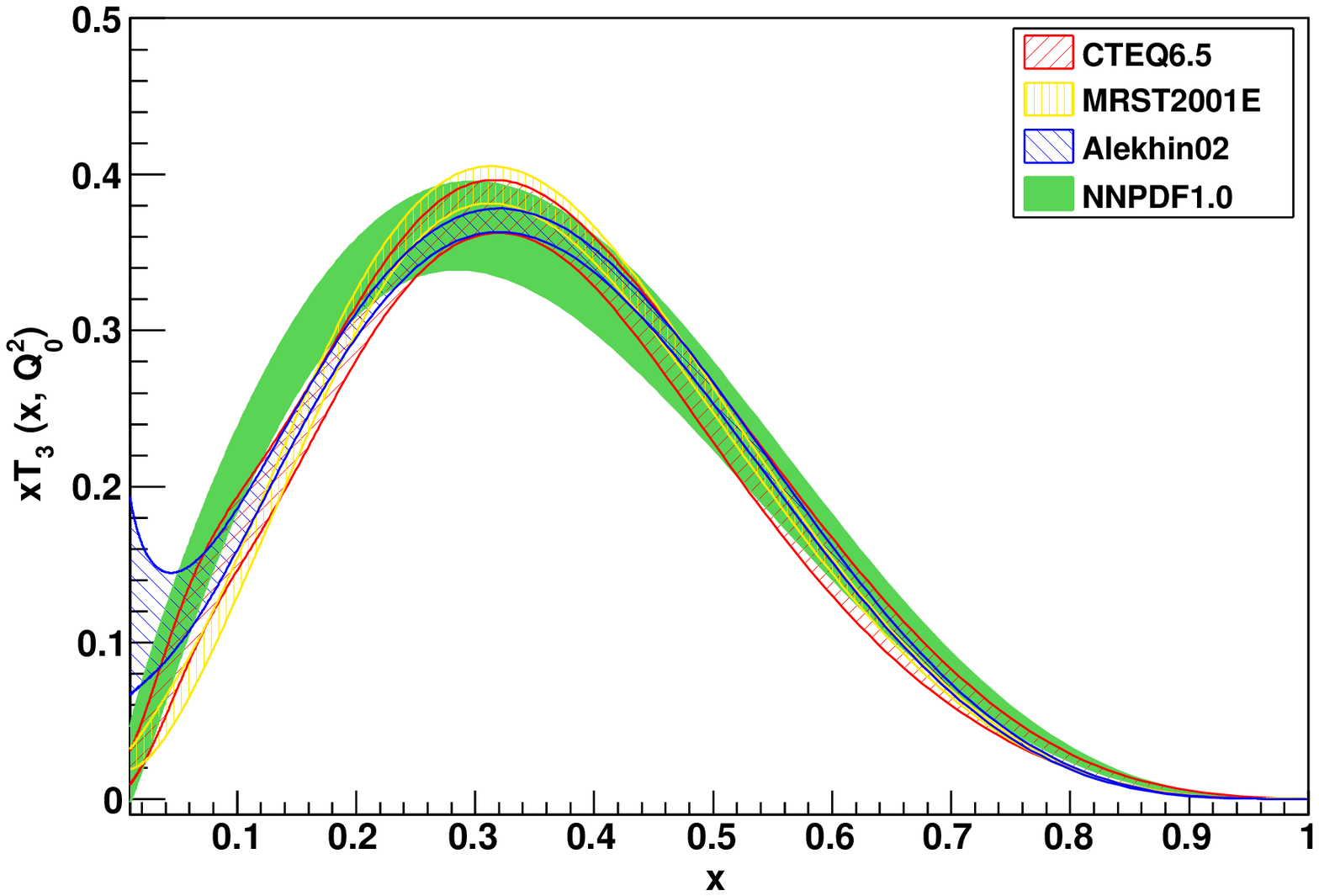}
\includegraphics[width=6cm]{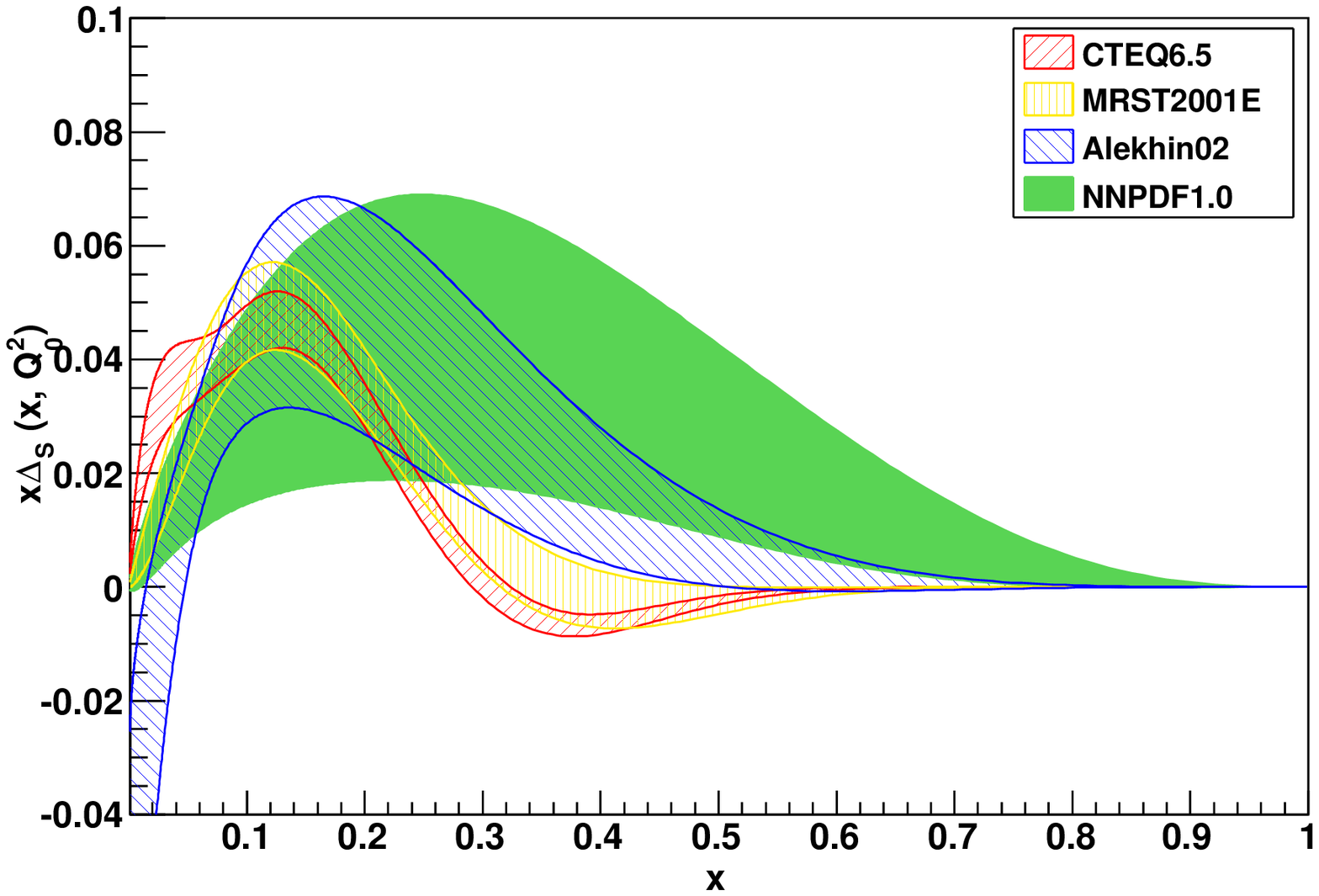}
\caption{The valence, triplet and sea asymmetry PDF at the initial scale $Q_0^2=2$ GeV$^2$}
\label{fig:vts}
\end{figure}
The five parton distributions which constitute our basis set are displayed in Fig.~\ref{fig:sg} and \ref{fig:vts} as a function of x at the fixed initial scale. Our results compared to those of the most recent NLO parton sets \cite{MRST,CTEQ,alekhin2} are in reasonable agreement especially in the data region. Uncertainties of PDFs tend to be larger in the region where no data are available, while in the data region they tend to be generally little larger, in some cases comparable or even smaller. Note that the uncertainty bands are found without introducing any tolerance criterion, which would correspond to an upward rescaling of all experimental uncertainties.
\begin{figure}[htb]
\vspace{-0.7cm}
\includegraphics[width=6cm]{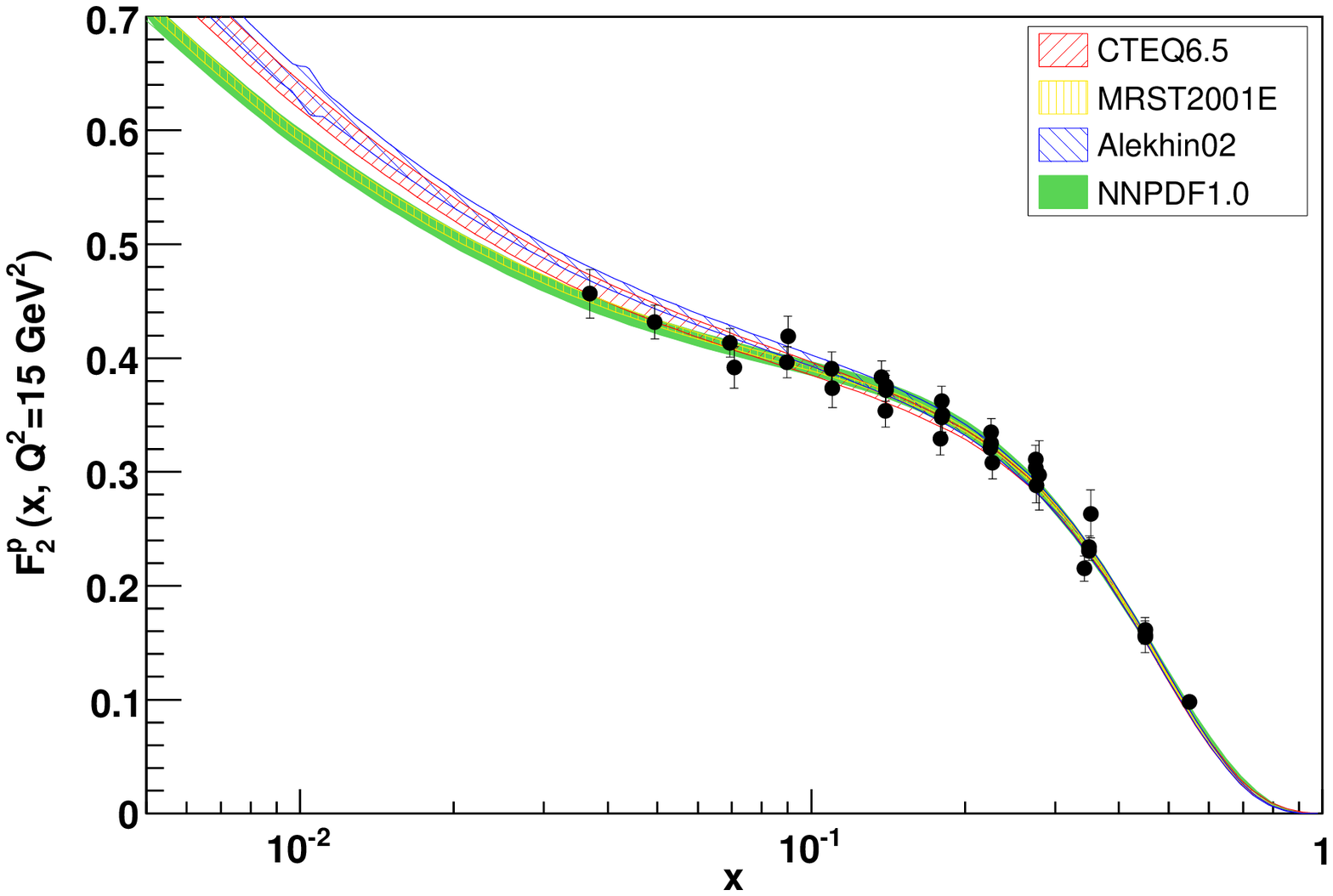}
\includegraphics[width=6cm]{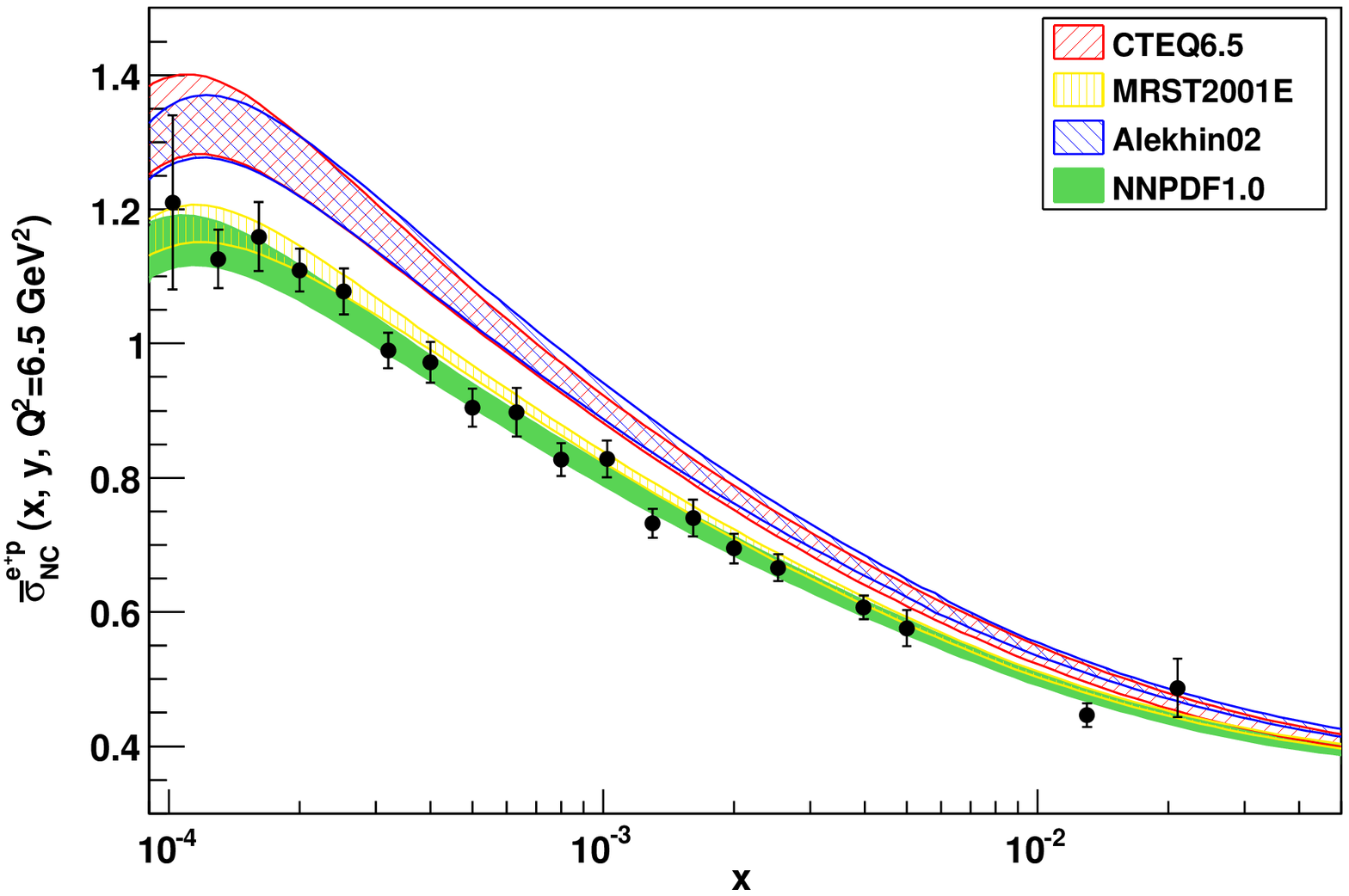}
\caption{Comparison of NLO theoretical predictions and data for one of the observables included in the fit. $F_2^p$ at $Q^2=$ 15 GeV$^2$(top) and NC reduced cross section at $Q^2=$ 6.5 GeV$^2$ and $\sqrt{s}=$ 301 GeV (bottom).}
\label{fig:obs}
\end{figure}

\begin{figure}[htb]
\vspace{-0.9cm}
\includegraphics[width=6cm]{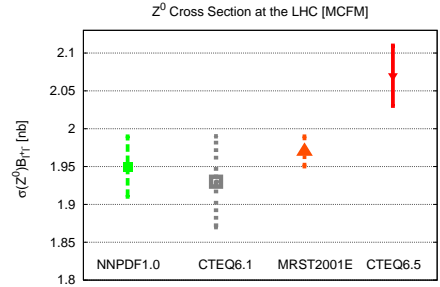}
\caption{Comparison of NLO theoretical predictions for the Z-production total cross section at the LHC.}
\label{fig:pred}
\end{figure}

In Fig.~\ref{fig:obs} the theoretical prediction obtained using NLO QCD and the NNPDF1.0 set
is compared to the data, for some representative deep-inelastic observables included in the fit. 
The differences between predictions obtained using different parton sets are
smaller for these observables than they are for the parton distributions themselves, as it should be
given that these data have been used in the determination of all the partonic sets with the exception of the
CHORUS data. In Fig.~\ref{fig:pred} we show as an illustrative example the total cross sections for the $Z$ production at the LHC. 
All cross sections have been computed at NLO using MCFM~\cite{mcfm},
using a sample of  $N_{\rm rep}=100$ replicas, which is fully adequate for this purpose. 
We find good agreement of central values with the CTEQ6.1
computation, as expected, given that it uses a ZM-VFN number scheme for heavy quarks as we do.

The NNPDF1.0 is the first full parton set based on this new approach and it is available in the LHAPDF interface~\cite{lhapdf}. It can be further improved in many respects. First of all a wider set of data besides DIS should be included as well as heavy quark thresholds should be treated more accurately. On aside, a set of NNLO parton distributions should be produced, both with the purpose of estimating uncertainties on the NLO results and also for some precision applications; large and small $x$ resummation corrections should also be considered.


\end{document}